# Spatially-resolved relaxation dynamics of photoinduced quasiparticles in underdoped $YBa_2Cu_3O_{7-\delta}$

C. W. Luo[1], P. T. Shih[1], Y.-J. Chen[2], M. H. Chen[1], K. H. Wu[1], J. Y. Juang[1], J.-Y. Lin[2], T. M. Uen[1] and Y. S. Gou[1]

[1] Department of Electrophysics, National Chiao Tung University, Hsinchu, Taiwan, R.O.C.

[2] Institute of physics, National Chiao Tung University, Hsinchu, Taiwan, R.O.C.

The spatially-resolved relaxation characteristics of photoinduced quasiparticles (QPs) in $CuO_2$ planes of underdoped YBCO are disclosed by polarized fs time-resolved spectroscopy. The relaxation time ($\tau$) along *b* axis diverges at $T_c$, and appears to be governed by a temperature-dependent gap $\Delta(T)$ at $T < T_c$. Furthermore, for $T > T_c$, a monotonic increase of $\tau$ with decreasing $T$ along the *b* axis and *ab* diagonal was observed and can be attributed to a temperature-independent gap $\Delta_p$. The results lend support to recombination dominant scenario of QP dynamics. However, the QP thermalization may take part along the nodal direction in the highly underdoped samples.

Recently, the dominant mechanism governing the recovery dynamics of photoinduced nonequilibrium quasiparticles (QPs) in YBa$_2$Cu$_3$O$_{7-\delta}$ (YBCO) high-$T_c$ superconductors has generated intense discussions [1-4]. Segre *et al.* [1] measured the temperature ($T$) and excitation energy density dependence of the photoinduced change of reflectivity in single crystals of YBa$_2$Cu$_3$O$_{6.5}$ (Ortho II), and proposed that the relaxation of photoinduced reflectivity is due to the dynamics of QP thermalization rather than their recombination into condensates. Subsequently, Gedik *et al.* [2] argued based on the results of reference [1] that the laser induced QP primarily occupy states near the antinodal regions of the Brillouin zone, and that the divergence of the relaxation time as $T \rightarrow 0$ is a consequence of momentum and energy conservation in electron-electron scattering. In a Comment to Ref. [1], Demsar *et al.* [3] argued that the decay rate of reflectivity change ($\Delta R/R$) seen by the probe pulses is a result of excited state absorption from the photoexcited QPs, and the recovery dynamics is governed by QP recombination. Similar quasiparticle lifetime results have been used to infer gap formation in both cuprate superconductors [5, 6] and systems exhibiting charge density waves [7, 8]. Since the interpretation of the QP recombination is at the heart of using femtosecond pump-probe spectroscopy to extract the magnitudes [5, 6] and spatial symmetries [9] of both the superconducting gap and pseudogap, the clarification of the issue under discussion is of essential importance. As suggested by Segre *et al.* [1], one way of resolving the above dispute could stem from coordinate measurements of photoinduced responses. In this paper, through polarized femtosecond pump-probe spectroscopy, *assuming two energy scales (gaps) existing in the samples*, we present a detailed analysis of the orientation-dependence of QP relaxation in a series of purely oriented YBCO films with different underdoping levels. The spatially-resolved QP relaxation dynamics had revealed significant anisotropy in ΔR/R along various crystalline axes [9] indicating that the above-mentioned discrepancies may be reconciled if the relaxation dynamics is revealed in respective channels.

In this study, highly in–plane aligned (100)- and (110)-oriented YBCO thin films prepared by pulsed laser deposition were used. The detailed growth conditions and structure-property characterizations of the films were reported elsewhere [10-12]. The percentage of the in-plane alignment for both (100) and (110) films was larger than 97% as determined by the x-ray Φ-scanning. In order to distinguish the directions of the axes in the thin film plane, the polarized x-ray absorption near-edge spectroscopy (XANES) of the O $K$-edge was carried out and successfully identified the $a$-, $b$-, and $c$-axes of YBCO thin films [10, 11]. In addition, we used the encapsulated bulk annealing method [13] to manipulate the oxygen content of the YBCO films. Although the oxygen content of the films can only be estimated from the corresponding $T_c$ obtained, we emphasize that this method is capable of controlling the oxygen content of the YBCO films precisely and reversibly. Furthermore, by using this method, all the measurements with various oxygen deficiencies can be performed on *a single YBCO thin film*. As a result, any changes in the superconducting properties should arise mainly from the effects of the oxygen content. The possible complications originated from individual film structures are minimized.

We measured the orientation-dependent ΔR/R at a photon energy of 1.5 eV. The change of ΔR/R is assumed to originate from the subsequent relaxation dynamics of the QPs excited by the pumping laser with the same photon energy. The details of the polarized pump-probe scheme have been described previously [9]. Briefly, the optical pulses were produced by a mode-locked Ti:sapphire laser with a 75 MHz train of 20 fs pulses. The ratio between the average power of the pump and probe beams was set at 40:1. The typical energy density of the pump pulses was ~ 4.4 μJ/cm$^2$, and the pulses were modulated at 87 KHz with an AO modulator. The reflected signals, typically very small, were detected using a lock-in amplifier. As discussed in detail previously [9], the measurements of ΔR/R along the *b* axis and the *ab* diagonal (oriented at 45° relative to the *a* or *b* axis in the *ab* plane) could be carried out with the availability of (100)- and (110)-YBCO thin films.

Fig. 1 shows the typical ΔR/R in the superconducting state ($T$ = 50 K) obtained by setting the polarizations of both of the pump and probe pulses parallel to the *b* axis or *ab* diagonal directions for films at nearly optimal doping (i.e. $T_c$ > 90 K). There are marked differences between these two major orientations not only in the amplitude [14], but also in the characteristic of relaxation. In general, ΔR/R in the superconducting state should be fitted by a double exponential decay function. One time scale is subpicosecond which relates to the electron-electron scattering, the other is the picosecond time scale which relates to the quasiparticles relaxation and to be discussed in this paper. Thus, the relaxation time $\tau$ of photoinduced QPs along various orientations could be extracted from ΔR/R, respectively. As plotted with solid squares in Fig. 2(a), the temperature-dependent relaxation time along the *b* axis in nearly optimal doped YBCO diverges when $T$ approaches $T_c$ from the low temperature side. This diverging behavior of temperature-dependent relaxation time at $T \approx T_c$ is believed to be a generic manifestation of superconducting gap opening. Indeed, consistent with most of other experimental data obtained from c-axis oriented films [5, 6], the current data obtained from the relaxation dynamics along b-axis can also be satisfactorily described by the theory of Kabanov *et al.* [5] as shown by the solid line in Fig. 2(a). The relevant parameters used in the $\tau(T)$ expression for QP relaxation (Eq. (1) below)

$$\tau(T) = \frac{\hbar\omega^2 \ln\left[\dfrac{1}{\varepsilon_I/2N(0)\Delta(0)^2 + \exp(-\Delta(T)/k_BT)}\right]}{12\Gamma_\omega \Delta(T)^2} \quad (1)$$

are $\Gamma_\omega \approx 24 \pm 4$ cm$^{-1}$ for the Raman phonon linewidth of $A_{1g}$-symmetry apical O(4) phonon mode (i.e. phonon frequency $\omega \approx 500$ cm$^{-1}$) in YBCO – which has been shown to be particularly anharmonic [15], $\varepsilon_I \approx 20\times10^{-20}$ J cell$^{-1}$ for the energy density per unit cell deposited by the incident laser pulse, $N(0) = 2.2 \sim 5$ eV$^{-1}$cell$^{-1}$spin$^{-1}$ for the density of state at $E_F$ [16], and $\Delta(T) = \Delta(0)[1-(T/T_c)]^{0.5 \pm 0.14}$ being a temperature-dependent function with $\Delta(0) \sim 454$ K, respectively. In contrast, for data measured along the *ab* diagonal the behavior of $\tau(T)$ is markedly different. As illustrated by the solid squares shown in Fig. 3, $\tau$ increases monotonically with decreasing temperature. In this case, the data though can also be described by a similar equation [17] (Eq. (2) shown below), the temperature-dependent gap $\Delta(T)$ in Eq. (1), however, has to be replaced by a temperature-independent gap $\Delta_p$. The fit to the data (dashed lines in Fig. 3) was obtained by setting $\Delta_p \sim 315$ K with other parameters fixed.

$$\tau(T) = \frac{\hbar\omega^2 \ln\left[\dfrac{1}{\varepsilon_I/2N(0)\Delta_p^2 + \exp(-\Delta_p/k_BT)}\right]}{12\Gamma_\omega \Delta_p^2} \quad (2)$$

These results indicate that the relaxation dynamics of QPs in the *ab* plane is strongly anisotropic. Namely, the QP relaxation channel through the *b* axis and the *ab* diagonal are respectively dominated by a temperature-dependent superconducting gap and a temperature-independent gap which could be associated with the pseudogap in Ref. 5 (Further discussions on $\Delta(T)$ and $\Delta_p$ are presented later in this paper.) Furthermore, it is noted that, by comparing the amplitude (i.e. the number of QP) and the relaxation time scale of ΔR/R along the two axes, the overall results in the *ab* planes will be dominated by the responses from *b* axis or *a*-axis (but it is inaccessible in our (100)-films), if the measurement is not properly resolved spatially. In that case, the information carried by the nodal QPs will be largely overlooked and we believe it might be what has been encountered by many early reports when dealing with (001)-YBCO thin films [5, 6, 18, 19].

When the film becomes more underdoped ($T_c$ = 55.7 K), the divergence of $\tau(T)$ near $T_c$ along the *b* axis is smeared and the overall relaxation time is shortened as demonstrated by the open circles in Fig. 2(a). To further compare $\tau(T)$ of the optimally doped and the underdoped samples, Fig. 2(b) shows the normalized $\tau(T)$ vs. the reduced temperature $T/T_c$, where $\tau(T)$ is normalized to the longest one observed above $T_c$ for each sample. It is noted note that $\tau(T)$ increases gradually with decreasing temperature for $T > T_c$ (or $T/T_c > 1$) for the underdoped sample. This is different from that in the optimal-doping case (solid squares in Fig. 2(b)), where $\tau(T)$ remains near constant for $T > 1.5T_c$. The behavior of $\tau(T)$ for $T > T_c$ in this case, in fact, appears to be more similar to that found along the *ab* diagonal [solid squares in Fig. 3]. More quantitatively, by fitting the data at $T < T_c$ (Fig. 2(b)) with Eq. (1), the superconducting gap at zero temperature $\Delta(0) \approx 560$ K is obtained, and the fitting with Eq. (2) [dashed line in Fig. 2(b)] above $T_c$ gives $\Delta_p \sim 301$ K for this underdoped sample. and demonstrates the contrast *T* dependences of $\tau(T)$ in the optimally doped and underdoped samples, respectively. The above analyses thus suggest that, for more underdoped samples, the recombination-induced relaxation dynamics of QPs along *b*-axis may have been governed by the simultaneous existence of two gaps. Furthermore, the normalized $\tau(T)$ of the optimally doped sample shoots up around $T_c$ by one order of magnitude higher than that of the underdoped sample. This difference appears to be genuine and can not be attributed to any thermal smearing, for $T_c$ of the former is actually higher. Overall, the data in Fig. 2 clearly demonstrate the contrast $T$ dependences of $\tau(T)$ in the optimally doped and underdoped samples, respectively.

Now turn attention to what happened to the *ab* diagonal when the sample becomes underdoped. As shown by the open circles in Fig. 3, $\tau(T)$ along the *ab* diagonal, though still exhibiting the gradual change fashion, appears to increase with underdoping at all temperatures. Furthermore, the increasing rate at low temperatures also becomes steeper for the more underdoped case. The fit with Eq. (2) [dotted line in Fig. 3] in this case gives $\Delta_p \sim 170$ K, significantly reduced from $\Delta_p \sim 315$ K for the optimal-doped case. The slower relaxation of the QPs residing around the nodal direction may thus simply reflect the shrinking of $\Delta_p$ along the *ab* diagonal when the sample becomes underdoped. Nonetheless, due to the shrinking of $\Delta_p$, it is natural to ask whether the relaxation is still of recombination-dominant characteristic or it may have shifted to other process, such as the thermalization by QP-pair scattering suggested by Segre *et al.* [1] and Gedik *et al.* [2, 4]. To explore this alternative, we plot in Fig. 4 the QP decay rate ($1/\tau$) along the *ab* diagonal as a function of the reduced temperature together with some data taken from Ref. [1]. The $1/\tau(T)$ behavior of the nodal QPs for the optimal-doping case is nowhere close to the $(T/T_c)^3$ dependence, and can be solely described by the recombination process of Eq. (2) (dashed line in Fig. 4) dominated by a relatively large $\Delta_p$ as we mentioned previously. It is also evident that with further underdoping $1/\tau(T)$ along the *ab* diagonal gradually shifts away from the behavior of QP recombination which is manifested by the slope of the dashed line in Fig. 4. Actually, it exhibits a tendency of following the $(T/T_c)^3$-dependence (the solid line in Fig. 4), displaying the characteristic of scattering-induced thermalization [20] for $T < T_c$.

From the above results, it is apparent that the relaxation dynamics of QPs in $CuO_2$ (or *ab*) plane of YBCO is anisotropic and doping-dependent. For optimal doping, the relaxation channels along the *b* axis and the *ab* diagonal were dominantly of recombination nature, and were respectively governed by the superconducting gap $\Delta(T)$ and pseudogap $\Delta_p$. This is consistent with the scenario provided by several literatures [3, 5, 6, 9]. However, as revealed by the spatially-resolved pump-probe measurements, owing to the shrinking of $\Delta_p$ along the *ab* diagonal with underdoping, mechanisms such as scattering-induced thermalization may take part in the relaxation processes of QPs. Both effects account for the strong divergence of relaxation time at low temperatures even under the experimental condition of nonzero pumping power. These relaxation channels may be easily overlooked in measurements that do not have spatial resolution to delineate the marked changes of $\Delta_p$ along nodal and antinodal directions in underdoped YBCO. On the other hand, the observation of pair-scattering thermalization of nodal QPs in underdoped YBCO reported by Segre *et al*. [1] can also be explained without invoking specific mechanisms [2]. Although in their experiments very low pumping power was used to minimize the effect of recombination via superconducting gap $\Delta(T)$ and $\Delta_p$ along antinodal orientation [1] as shown with the crosses and open triangles in Fig. 4, it is argued that the rapid increase in relaxation time may be simply due to the reduction of $\Delta_p$ along the nodal orientation as the sample becomes underdoped.

The doping dependence of $\Delta(T)$ and $\Delta_p$ are intriguing. It should be noted that no evidence for the change of the pseudogap symmetry with doping was provided by ARPES results [21]. Nevertheless, the point is that the symmetry of $\Delta_p$ seems to be changing with hole-doping (i.e. the $\Delta_p$ displays a $d_{xy}$-spatial symmetry near the optimal-doping and the symmetry gradually evolves into a $d_{x^2-y^2}$-symmetry with decreasing the hole concentration of YBCO) *if* the experimental data are analyzed by Kabanov's model, which has been adopted in interpreting some of the time-resolved spectroscopy experiments [5, 6]. This discrepancy may simply arise from the fact that ARPES gives rather direct information about the quasiparticle spectrum while the pump-probe spectroscopy provides relatively indirect spectrum. Namely, the spectrum is convolved with itself as well as relevant inelastic scattering processes such as phonons, magnons, and so on. Our paper, thus, either provides one possible scenario of identifying the symmetry of the highly debated high-energy pseudogap and reveals how it evolves with hole-doping *or* a complete theory for interpreting the time-resolved spectroscopy experiments has not yet been available up to now. The validity of attributing the obtained experimental results to the symmetry change of high-energy pseudogap should be judged carefully by further experiments and more developed theories.

In summary, we have demonstrated that the photoinduced QP relaxation dynamics in two major crystalline directions could be clearly revealed by the polarized femtosecond time-resolved spectroscopy combined with the well characterized (100)- and (110)-oriented YBCO thin films. The results suggest that there are two different energy scales, i.e. the superconducting gap $\Delta(T)$ and the temperature independent gap $\Delta_p$, that may have played important roles in governing the QP relaxation dynamics of cuprate superconductors. Moreover, the thermalization by QP-pair scattering around the nodal regions may also take part in the relaxation process when $\Delta_p$ along the nodal direction decreases with increasing underdoping.

This work was supported by the National Science Council of Taiwan, R.O.C. under grant: NSC92-2112-M-009-031.

**Figure Captions:**

Fig. 1. The temperature dependence of ΔR/R at 50 K for the optimal doping along the *b* axis and the *ab* diagonal, respectively. The solid lines are the exponential decay fits.

Fig. 2. The temperature-dependent relaxation time of ΔR/R along the *b* axis at various oxygen contents obtained by fitting such as in Fig. 1. (b) The normalized relaxation time in (a) is depicted as the normalized relaxation time vs. the reduced temperature $T/T_c$. The relaxation time is normalized to the longest one observed above $T_c$ for each sample. The vertical axis of the inset is rescaled to show the different magnitudes of the normalized relaxation time in these two samples. The solid and dashed lines are fitted, respectively, by Eq. (1) and Eq. (2).

Fig. 3. The temperature-dependent relaxation time of ΔR/R along the *ab* diagonal at various oxygen contents obtained by fitting such as in Fig. 1. The dashed and dotted lines are fitted by Eq. (2).

Fig. 4. The decay rates (1/τ) of QPs along ab diagonal are plotted as a function of the reduced temperature for various $T_c$'s. The solid squares and open circles were directly taken from Fig. 2(a). The crosses and open triangles were taken from Ref. 1, which were measured, respectively, under 0 and ~ 3.3 μJ/cm$^2$ pump intensity in an untwinned single crystal of YBa$_2$Cu$_3$O$_{6.5}$ ($T_c$ = 57 K) with the ortho II structure. The dashed line is from Eq. (2).

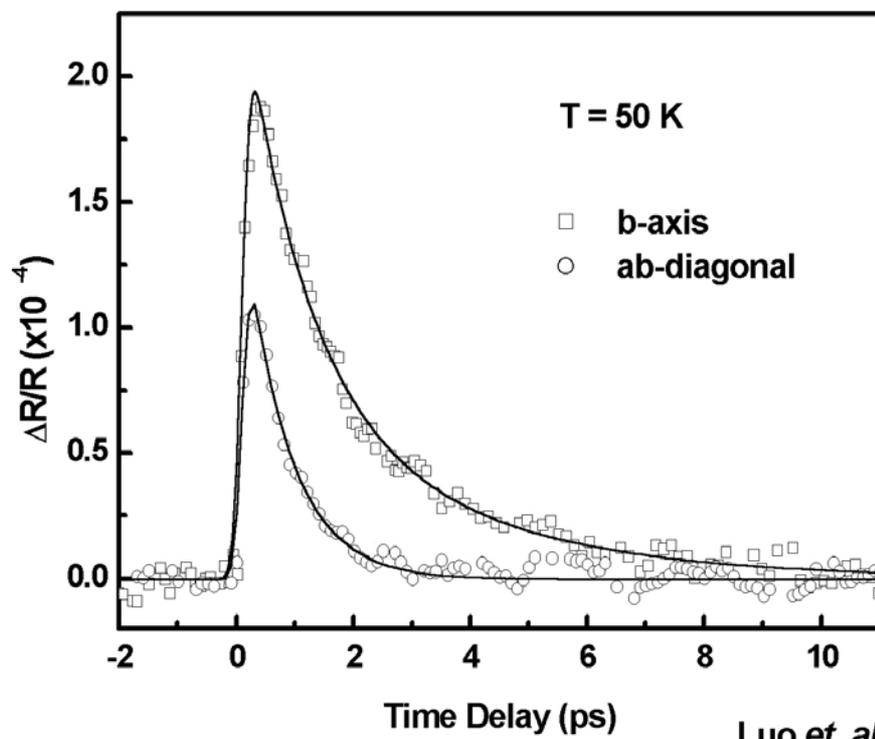

Luo *et. al.* --- Fig. 1

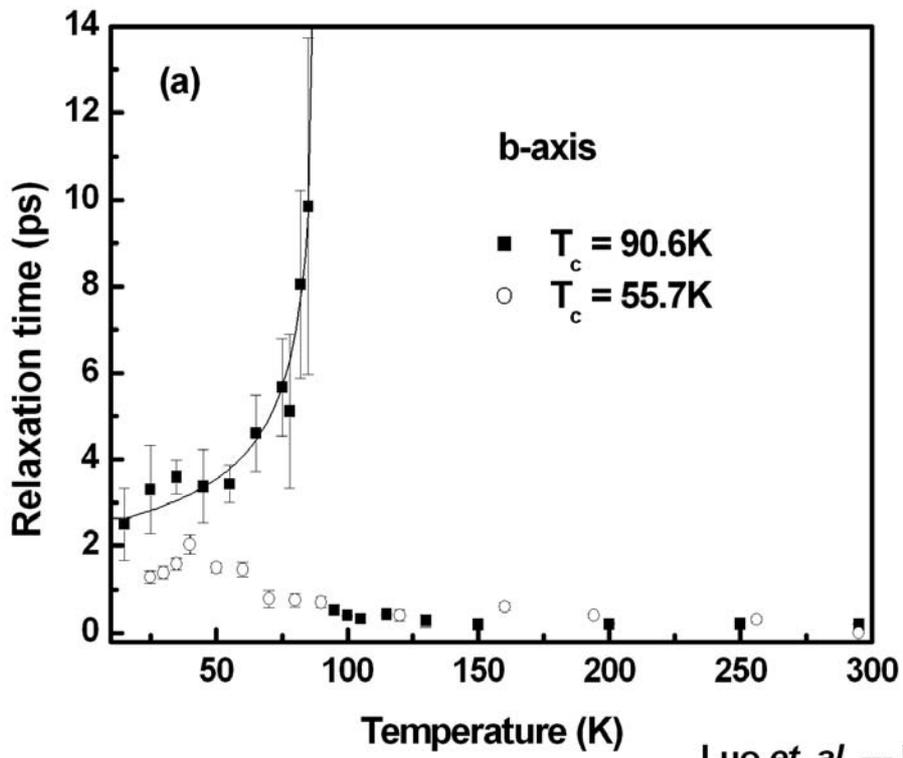

Luo *et. al.* --- Fig. 2(a)

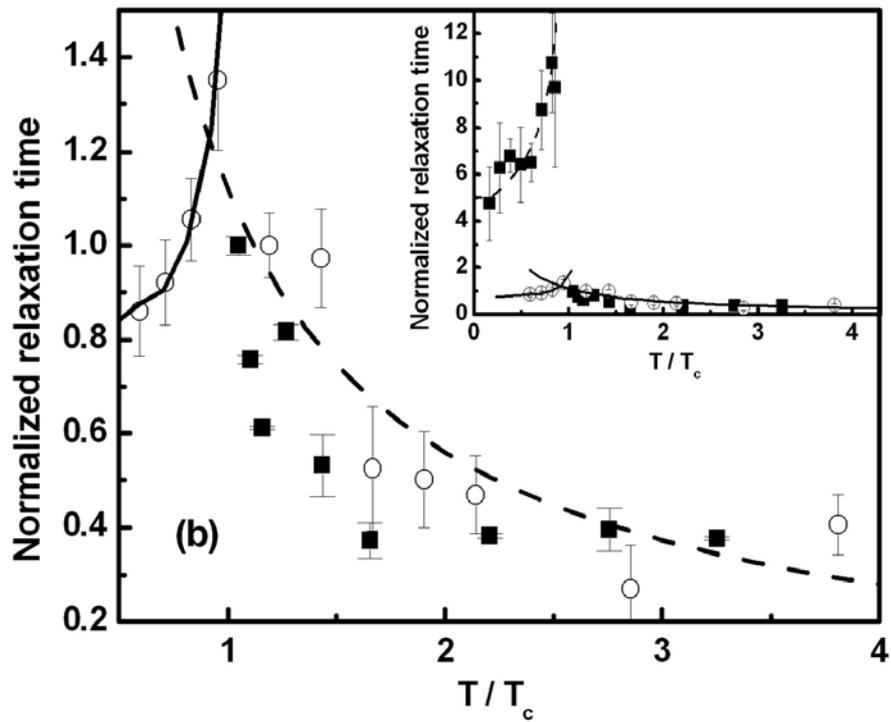

Luo *et. al.* --- Fig. 2(b)

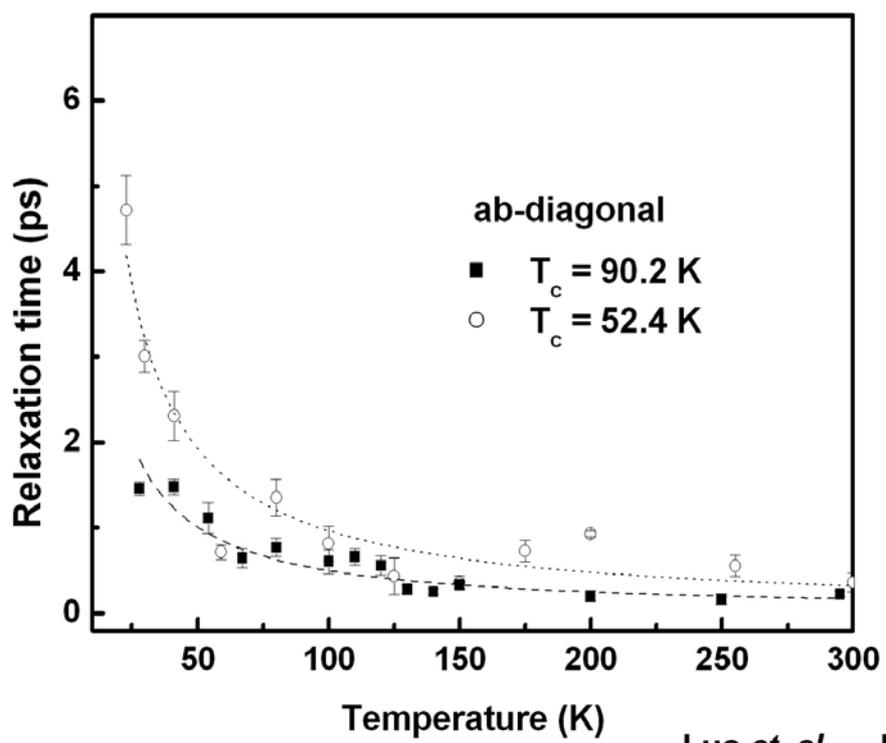

Luo *et. al.* --- Fig. 3

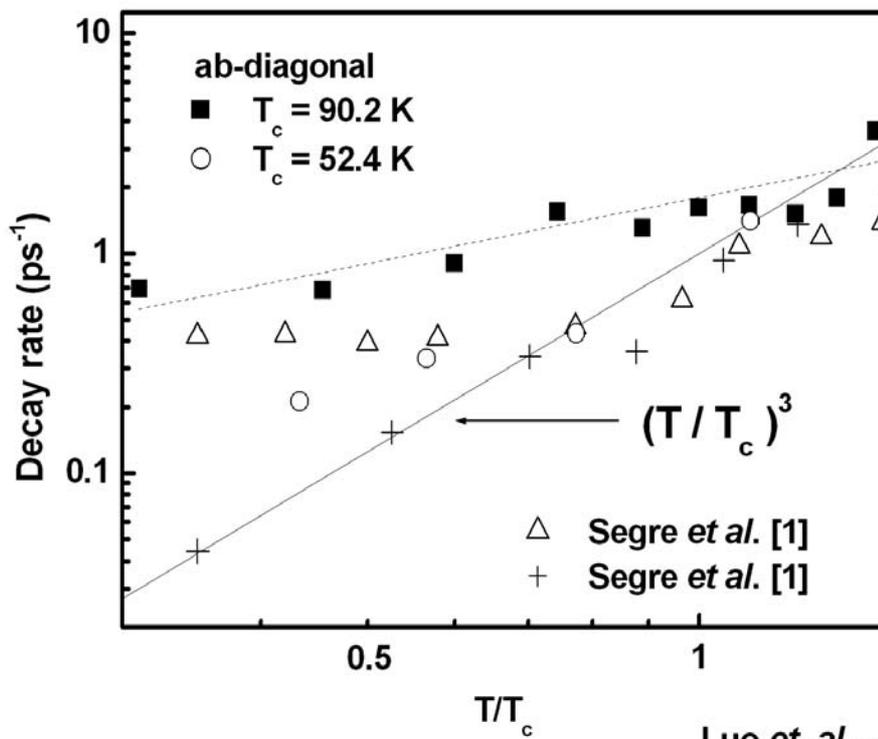

Luo *et. al.* --- Fig. 4